\documentclass{aa}
\usepackage{graphicx,epsfig,amssymb}

\newcommand{\los}{{\em los}}
\newcommand{\Mlim}{M_{\rm lim}}
\newcommand{\Ylim}{Y_{\rm lim}}
\newcommand{\xidm}{\xi_{\rm dm}}
\newcommand{\fgas}{f_{\rm gas}}
\newcommand{\Csz}{C^{\rm sz}}
\newcommand{\Ho}{H_{\rm o}}
\newcommand{\OmM}{\Omega_{\rm M}}
\newcommand{\OmB}{\Omega_{\rm B}}
\newcommand{\OmL}{\Omega_\Lambda}
\newcommand{\Da}{D_{\rm ang}}
\newcommand{\da}{d_{\rm ang}}
\newcommand{\npair}{n_{\rm pair}}
\newcommand{\Dtheta}{\Delta\theta}

\newcommand{\dfig}[6]{\centerline{\epsfig{file=#1, width=#2\linewidth, angle=#3} \epsfig{file=#4, width=#5\linewidth, angle=#6}}}

\begin{document}
   \title{On the Angular Correlation Function of SZ Clusters : \\
          Extracting cosmological information from a 2D catalog}

   \subtitle{}

\titlerunning{Cosmological information from a 2D SZ catalog}

   \author{S. Mei
          \inst{1}
          \and
          J.G. Bartlett \inst{2}
%\fnmsep\thanks{Just to show the usage
%          of the elements in the author field}
          }

   \offprints{S. Mei}

   \institute{Institut d'Astrophysique Spatiale, Orsay, France\\
              \email{simona.mei@ias.u-psud.fr}
         \and
             APC -- Universit{\'e} Paris 7, Paris, France\\
             \email{bartlett@cdf.in2p3.fr}
%             \thanks{The university of heaven temporarily does not
%                     accept e-mails}
             }

   \date{Received February 25th; Accepted August 19th}

   \abstract{We discuss the angular correlation function 
of Sunyaev--Zel'dovich (SZ)--detected galaxy clusters
as a cosmological probe.  As a projection of 
the real--space cluster correlation function, the angular
function samples the underlying SZ catalog redshift distribution.
It offers a way to study cosmology and cluster evolution
directly with the two--dimensional catalog, even before 
extensive follow--up observations, thereby facilitating
the immediate scientific
return from SZ surveys.  As a simple illustration of the
information content of the angular function, we examine its
dependence on the parameter pair $(\OmM, \sigma_8)$ in flat 
cosmologies.  We discuss sources of modeling uncertainty and
consider application to the future Planck SZ catalog, 
showing how these two parameters and the normalization
of the SZ flux--mass relation can be simultaneously found when the 
local X--ray cluster abundance constraint is included.
   }

   \maketitle
%
%________________________________________________________________

\section{Introduction}

The Sunyaev--Zel'dovich (SZ) effect 
(Sunyaev \& Zel'dovich 1970, 1972)
has become a practical observational tool for studying 
galaxy clusters and cosmology (for recent reviews see
Birkinshaw 1999; Carlstrom et al. 2002).  
Current observations of individual clusters, when combined with
X--ray observations, constrain cosmological parameters
via gas mass fractions (Grego et al. 2002) and angular--diameter 
distance determinations (Mason et al. 2001; Jones et al. 2001; 
Reese et al. 2002).  Multi--band millimeter observations
of a handful of clusters have already been used to set limits
on peculiar velocities (Holzapfel et al. 1997; 
Benson et al. 2003), and theoretical studies of this technique show its 
promise for the future (Aghanim et al. 2002a, 2002b; 
Holder 2002).
A new generation of optimized, dedicated 
instruments, both large--format bolometer arrays 
and interferometers with high sensitivity receivers, 
will qualitatively improve these studies.
And the arrival of these instruments within the next few years,
in addition to the Planck 
mission\footnote{{\tt http://astro.estec.esa.nl/SA-general/Projects/Planck/}},
will move the field forward to its next important step: 
surveying.  This will open a new observational
window onto large--scale structure and its evolution 
out to large redshifts.

The ultimate goal of these SZ surveys is the construction
of large cluster catalogs with multi--wavelength follow--up 
observations in order to perform cosmological studies; for
example, constraining cosmological parameters with 
the counts and redshift distribution (e.g., Barbosa et al. 1996; 
Eke et al. 1996; Colafrancesco et al. 1997; Haiman et al. 2001;
Holder et al. 2001; Kneissl et al. 2001; Weller et al. 2002; 
Benson et al. 2002).  Driving this effort are the
particular advantages of SZ--based cluster catalogs (Bartlett 2000;
Bartlett 2001):  Firstly, SZ surveys are intrinsically efficient
at finding clusters at large redshift, due to the surface brightness
constancy of the SZ effect\footnote{for a cluster of fixed properties}.
The thermal SZ spectrum is furthermore universal, the same for all
clusters at any redshift\footnote{in the non--relativistic limit}.
Other emission mechanisms, in contrast, suffer from cosmological 
dimming and the need for accurate $k$--corrections.  Secondly,
SZ surveys select clusters based on their thermal energy.
Since the spectrum is the same for all clusters, the total observable
SZ flux from a cluster can be expressed in a frequency independent
manner as the integrated Compton $y$--parameter, 
$Y=\int d\Omega\; y(\hat{n})$, where the integral is over 
the cluster profile (see Equation \ref{eq:Yscaling} below).  
The $y$ parameter being the pressure integrated along the
line--of--sight ($y\propto \int dl\; nT)$, this then implies 
$Y\propto MT$, i.e., the thermal energy of the 
intracluster medium (ICM).  This
is important, because the total thermal energy of the ICM
is given by energy re-partition during cluster collapse and
is independent of any thermal or spatial structure in the gas.
It is hence a more robust quantity than, for example, 
the X--ray emission measure that depends in a more complicated
fashion on both the ICM density and temperature.  Hydrodynamical
simulations confirm this expectation by showing a tight
SZ flux--mass relationship with little scatter (da Silva 
et al. 2003).  Object selection
in a flux--limited SZ catalog is therefore relatively easy
to interpret in terms of cluster mass and redshift.
For instance, it is easy to show that the minimum
detectable cluster mass is almost independent of redshift.
This is particularly advantageous for evolutionary studies,
because one is able to follow the evolution of the same
kind of object over redshift, instead of comparing
massive objects at high redshift to less massive ones at
low redshift, as is the case with X--ray samples.

Detailed follow--up of SZ surveys will, however, be time--consuming, 
and an enormous effort for the more than $10^4$ clusters expected 
from Planck. Large--area photometric surveys in the optical
and infrared (e.g., SDSS) will help (Bartelmann \& White 2002), 
but it is important to identify the kind of science 
that may be done directly with the two dimensional catalog of
cluster positions and SZ fluxes, what we will refer to in the
following as the SZ photometric catalog.  
This will certainly be the first science to be performed.  
Source counts represent the primary avenue of 2D study 
that has been discussed extensively in the literature.
In this {\em paper}, we examine the next higher order
catalog statistic, namely, the {\em angular} correlation 
function $w(\theta)$ of SZ--detected clusters (Diaferio 
et al. 2003).  We quantify 
its information content and study its potential use as a 
cosmological probe.  The angular function samples the catalog 
redshift distribution, because it is a projection of the
real--space correlation function along the line--of--sight.
With an appropriate model for the real--space correlation
function of the catalog, we may gain some insight on this
distribution, and hence on the underlying cosmological model.

Moscardini et al. (2002) studied the 3D clustering 
properties of SZ detected clusters, taking the Planck 
survey as an example and accounting for evolutionary
effects along the past light--cone.
Diaferio et al. (2003) examined the angular function
of SZ clusters as a means of identifying probable
physical cluster pairs (in 3D) and superclusters.
Our modeling is similar to theirs, although we 
focus instead on the angular function as a cosmological
probe permitting the extraction of cosmological information
from a photometric SZ catalog.  
The idea is of course not new, and has been applied in 
the past to, for example, optical and X--ray cluster catalogs;
but we reiterate the advantages of an SZ catalog 
in this context:  the cluster selection function
is relatively easy to model (compared to other observing
bands), and it extends out to large redshift, giving a longer
base--line for viewing evolutionary effects.

We should distinguish at the outset the difference
between the angular power spectrum, $\Csz_l$, of SZ--induced 
temperature fluctuations (secondary anisotropies in the cosmic 
microwave background [CMB]) and 
the angular correlation function of {\em detected clusters} in
a SZ survey, $w(\theta)$.  The angular power spectrum $\Csz_l$ 
is a two--point statistic quantifying the integrated contribution of
the entire cluster population to the CMB sky brightness fluctuations.
It is dominated by the Poisson term and its overall shape is
determined by the mean SZ profile of clusters.  Cluster--cluster 
correlations add additional power on the order of $20-30$\% 
of the pure Poisson term (Komatsu \& Kitayama 1999).
Since it is defined relative to the mean cosmic microwave 
background temperature, we expect the SZ fluctuation 
power to increase with the surface density of clusters on the sky.
This is quantitatively confirmed by both numerical simulations and 
analytical calculations that indicate $\Csz_l\propto \sigma_8^7$ 
(all other factors held constant), where $\sigma_8$ is the amplitude 
the density perturbations, the quantity most directly influencing 
cluster abundance.  The fluctuation power spectrum
is an analysis method appropriate in a low signal--to--noise 
context (the current situation) where individual source 
identification is not possible\footnote{Either for low signal--to--noise 
observations or when pushing constraints on source counts below the detection 
threshold.  It is the SZ equivalent of the ``background fluctuation 
analysis'' of radio and X--ray Astronomy.}.
Fluctuations induced by the SZ effect have been invoked 
as a possible explanation for the excess power at high multipole 
$l$ reported by the CBI collaboration (Bond et al. 2002) 
and consistent with new VSA data (Grainge et al. 2002).
If this were entirely due to the cluster population, 
it would imply a surprisingly large value for $\sigma_8$ 
($>1$; Bond et al. 2002; Komatsu \& Seljak 2002; 
Holder 2002), although an important contribution from
heated gas at reionization may also be expected (Oh et al. 2003).

The angular function $w(\theta)$ quantifies the 
projected clustering of a 2D catalog of individually detected 
clusters.  It refers to the object positions and makes no 
reference to the mean background sky brightness.  
There are several ways to imagine using the information
contained in the SZ cluster angular function (e.g., Diaferio et al. 2003).  In the following
we choose to illustrate its use by examining constraints
on the matter density $\OmM$ and $\sigma_8$ in the context of
flat CDM--like models. 
The SZ counts provide one constraint on a combination
of these parameters.  To extract additional information
from the 2D catalog using the angular function, we are forced 
to model the real--space cluster correlation function. 
In CDM scenarios, clusters form from peaks in the
density field whose clustering may be analytically calculated.
We adopt the approach proposed by Mo \& White (1996; 2002).  Any 
conclusions that we draw are, therefore, unavoidably dependent on this
clustering model (Moscardini et al. 2002; Diaferio et al. 2003); however, it is well founded 
in the context of CDM cosmogonies and compares well with the
results of numerical simulations, at least at redshifts lower than
$\sim 10$ (Reed et al. 2003; Jenkins et al. 2001).
Another important issue concerns the modeling of the 
intracluster medium.  Moscardini et al. (2002), for 
example, find non--negligible dependence of the 3D 
correlation function on ICM properties.  We discuss 
the issue below for the angular function, but we note 
here that it tends to be dominated by massive clusters 
(depending of course on the exact flux limit; see below)
that follow relatively well observed scaling laws.
As we show, adding constraints from the local
X--ray cluster abundance permits us to simultaneously
constrain the cosmological parameters and the cluster
baryon content.

In the following section, we give our master equation for 
calculating $w(\theta)$ and identify the necessary modeling
ingredients.  We outline our cluster model in Section 3. 
Results for the angular function are presented in Section 4, 
where we apply the results to constraining the cosmological 
parameters $\sigma_8$ and $\OmM$.  We discuss the 
influence of ICM physics on the results and how to use
additional information from the local cluster abundance
to simultaneously constrain this physics and the cosmology.
Our fiducial example is the Planck mission.  Section 5 closes
with a final discussion and summary.

\section{The Angular Correlation Function}
\label{sec:angcorr}

In this section we relate the angular correlation
function to the real--space correlation function
in the context of SZ observations. 
The 3--dimensional (auto)correlation function $\xi$  
quantifies the 2--point clustering of a population 
in terms of the probability in excess of Poisson of 
finding two objects at a separation $r$.
The angular correlation $w(\theta)$ of the same population is 
the projection of $\xi$ onto the sky:
\begin{eqnarray}
%\begin{equation}
d\Omega_1 d\Omega_2 \Sigma^2 [1 + w(\theta)] = \\
      d\Omega_1 d\Omega_2 \int d z_1 d z_2 \frac{dV_1}{d\Omega_1 dz_1} 
        \frac{dV_2}{d\Omega_2 dz_2} \times \nonumber \\
	\int dM_1dM_2 \frac{dn_1}{dM_1} 
        \frac{dn_2}{dM_2} (1 + \xi) \nonumber
\end{eqnarray}
%\end{equation}
where the integrals concern two lines--of--sight (\los) of solid
angles $d\Omega_1$ and $d\Omega_2$ separated on the sky by 
angle $\theta$.  In this expression, $dn/dM$ represents the
cluster mass function, and we assume the small angle 
approximation here and throughout. 
The surface density of sources (the counts) with integrated 
Compton parameter larger than $\Ylim$ (the SZ `flux' limit; see the
following section, Equation \ref{eq:Yscaling}) 
is given by 
\begin{equation}
\label{eq:counts}
\Sigma(\Ylim) = \int_0^\infty dz \frac{dV}{d\Omega dz} 
        \int_{\Mlim(\Ylim,z)}^\infty 
        dM \frac{dn}{dM} (M,z)
\end{equation}
The correlation function $\xi$ depends on cluster mass, redshift 
and, according to statistical isotropy, on physical 
separation $r(\theta)= r_1^2(z_1) + r_2^2(z_2) 
- 2 r_1 r_2 \cos(\theta)$, where $r_1$ and $r_2$ are 
angular--diameter distances.  Assuming that
correlations fall off sufficiently rapidly with distance, 
as is observed, 
we may take the two clusters to be at approximately 
the same redshift and write $\xi[M_1,M_2,z,r(\theta)]$.
We furthermore adopt a linear biasing scheme in which 
$\xi(M_1,M_2,z,r)=b(M_1,z)b(M_2,z)\xidm(z,r)$, where 
$\xidm(z,r)$ is the correlation function of the underlying 
cold dark matter and $b(M,z)$ is the bias factor
for clusters (see below).  Then, using the short--hand 
notation \mbox{$\phi(M,z,\Ylim)=(1/\Sigma)b(M,z)(dV/dzd\Omega)dn/dM$} 
for the joint 
distribution of clusters in mass and redshift, weighted by
the bias factor, we arrive at the expression 
\begin{eqnarray}
\label{eq:limber}
\lefteqn{w(\theta,\Ylim)  =} \nonumber\\
&&	 \int_0^\infty dz \int\int_{\Mlim(\Ylim,z)}^\infty
        dM_1 dM_2\; \phi (M_1,z) 
        \phi (M_2,z)  \times \nonumber\\
&&	\int dr \frac{dz}{dr}(z) 
        \xidm [z,r(\theta)] \nonumber\\ 
&&      \equiv \int_0^\infty dz  
        \Phi^2(z,\Ylim)\int dr 
        \frac{dz}{dr}(z) 
        \xidm [z,r(\theta)] \nonumber
\end{eqnarray}
a sort of Limber's equation appropriate for SZ sources
(Limber 1953; Peebles 1993; Diaferio et al. 2003) in which we explicitly
show the dependence on limiting flux $\Ylim$.  
From this equation we clearly see that the three key ingredients 
are the the mass--limit function $\Mlim(z,\Ylim)$, the distribution 
function $\phi$ and the correlation 
function $\xidm$.  We now discuss our modeling of each.

\section{The SZ Population}

The SZ cluster population inherits its properties from
two sources:  its constituent dark matter halos, 
whose properties are the sole result of gravitational 
evolution, and the relationship between observable SZ flux 
and these halos, governed by more difficult to model 
baryonic physics.  It is reasonable to characterize dark 
halos by their mass and redshift, and we will apply
the results of N--body simulations that give both their
abundance (i.e., the mass function) and spatial correlations 
(i.e., clustering bias $b(M,z)$ and the correlation
function $\xidm$) as a function of these two 
fundamental descriptors.  More difficult
to model, the baryonic physics of the cluster gas 
requires particular attention to various (and
at times contradictory) observational constraints
and theoretical scaling laws. 

\subsection{Halo properties}

The abundance of galaxy clusters is given by the mass function of
collapsed objects, which is completely specified once the linear
power spectrum of dark matter perturbations is specified.  
For the latter we adopt the BBKS (Bardeen et al. 1986) transfer function (see also below),
while for the mass function we employ the fitting formula (improved
Press--Schechter) given by Sheth \& Tormen (1999):
\begin{eqnarray}
\frac{dn}{dM}(M,z) dM = A\left(1 + \frac{1}{\nu'^{2 q}}\right) 
        \sqrt{\frac{2}{\pi}}
        \frac{\overline\rho}{M}\frac{d\nu'}{dM} \exp\left(\frac{-\nu'^2}
        {2}\right) dM
\end{eqnarray}
where $\overline{\rho}$ is the universal mean mass density and 
the constants $A \approx 0.322 $ and $q = 0.3$;  the parameter
$\nu' = \sqrt{a}\nu$, where 
$\nu \equiv \frac{\delta_c} {D(z)\sigma(M)}$ 
is the usual critical peak height ($\delta_c\approx 1.69$) 
normalized to the mass--density perturbation variance $\sigma(M)$ 
in spheres containing mass $M$, and the constant 
$a=0.707$.  The expression for the growth factor for flat models
with $\Lambda > 0$ is taken from Carroll et al. (1992):
\begin{eqnarray}
D(z,\OmM, \OmL)  = &\frac{g(z)}{g(0)(1+z)} \nonumber &\\
g(z,\OmM,\OmL)  \approx & \frac{5}{2} \OmM(z) [\OmM(z)^{4/7} - &\nonumber \\
- \OmL(z)& + (1 + \OmM(z)/2) (1+ \OmL(z)/70)]^{-1}\nonumber 
\end{eqnarray}
with the definitions $\OmM(z)\equiv \OmM(1+z)^3/E^2(z)$, 
$\OmL(z)\equiv \OmL/E^2(z)$,
and $E^2(z) = [\OmL + (1-\OmM-\OmL) (1+z)^2 + \OmM(1+z)^3]$;
$\Omega_M$ and $\Omega_\Lambda$ written without an explicit redshift
dependence will indicate present--day values ($z=0$).

As mentioned above, we use a linear bias scheme to relate
the cluster--cluster correlation function to that of the
dark matter and employ the analytic fitting formula 
for $b(M,z)$ given by Sheth et al. (2001);
the formula includes corrections for ellipsoidal perturbation
collapse:
\begin{eqnarray}
b(M,z) = 1 + \frac{1}{\sqrt{a}\delta_c(z)} \left[\sqrt{a} (a \nu^2) + 
\sqrt{a} b (a \nu^2)^{1-c}\right] - \nonumber \\
-\frac{(a \nu^2)^c}{(a \nu^2)^c + b (1-c)(1-c/2)}\nonumber
\end{eqnarray}
where $\delta_c$, $\nu$ and $a$ are given above, and 
$b = 0.5$ and $c = 0.6$. 
We model the linear dark matter perturbation spectrum with the 
BBKS transfer function with shape parameter {\em fixed at} $\Gamma=0.25$ 
and scale--invariant primordial density fluctuations ($n=1$).  
This seems to provide a good fit to galaxy clustering data 
(Percival et al. 2001) 
and is consistent with constraints on $n$ from CMB anisotropies (Spergel et al. 2003, and references therein). 
The resulting linear theory $\xidm$ is 
adequate on most scales ($\theta > 10 \arcmin$), 
although we also include non--linear 
corrections according to the fitting formula developed 
by Peacock \& Dodds (1996).

\begin{figure}
\includegraphics[width=8.5cm]{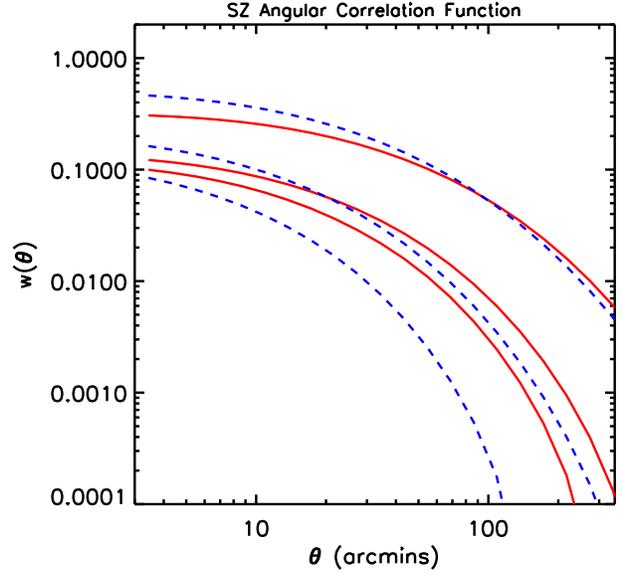}
\caption{Angular correlation function of SZ detected clusters
calculated in two different flat models ($\OmM=0.3$ and $\sigma_8 = 0.9$ as 
the dashed lines, and $\OmM=1$ and $\sigma_8 = 0.7$ as the solid lines) 
for three flux limits ($\Ylim=10^{-3}$, $10^{-4}$, $10^{-6}$~arcmin$^2$, 
decreasing from top to bottom).  
The models adopt the same linear matter power spectrum
(shape parameter $\Gamma=0.25$) and $Y_*=7.6\times 10^{-5} h^{7/6}$~arcmin$^2$
(see text).
}
\label{fig:models}
\end{figure}

\subsection{Intracluster Medium}
\label{sec:ICM}
With the abundance and clustering of  
halos now specified, we next relate the 
observable SZ flux to cluster mass and redshift.  
This relation is particularly robust from a theoretical 
viewpoint, contrary to, for example, the situation for X--ray 
luminosity. 
In the non--relativistic regime, the surface brightness of
the thermal SZ effect -- measured relative to the mean
sky intensity -- at position $\hat{\Omega}$ on a 
cluster image is 
\begin{displaymath}
\Delta i_\nu(\hat{\Omega}) = y(\hat{\Omega}) j_\nu
\end{displaymath}
where the Compton $y$ parameter is an integral of the pressure
along the line--of--sight%% ($\hat{\Omega}$)
\begin{displaymath}
y(\hat{\Omega}) = \int_{los\; \hat{\Omega}} dl \frac{kT}{mc^2}n\sigma_T 
\end{displaymath}
with $T$ the gas temperature (strictly speaking, that 
of the electrons), $\sigma_T$ the Thompson cross section,
$k$ and $m$ the Boltzmann constant and the electron mass,
respectively, and where $j_\nu$ is a universal spectral function that
is the same for all clusters, independent of their properties
(Birkinshaw 1999).
Since the thermal SZ spectrum is the same for all 
clusters, we may express the total flux in a frequency independent 
manner using the integrated Compton parameter $Y=\int d\Omega\; 
y(\hat{\Omega})$, where the integral is over the cluster profile.
The total flux density (e.g., in {\em Jy}) is then $S_\nu=Y j_\nu$.
This is the total flux (density) measured by an experiment with 
low angular resolution in which clusters are simply point 
sources. 
From these definitions, we find that the observable
flux is directly proportional to the total thermal energy of the ICM
(e.g., Barbosa et al. 1996):
\begin{equation}\label{eq:Yscaling}
Y(M,z) = \frac{k\sigma_T}{mc^2}\frac{N_{\rm e}T}{\Da^2(z)}
       \propto \frac{\fgas(M,z) T(M,z) M}{\Da^2(z)} 
\end{equation}
where $N_{\rm e}$ is the total number of electrons, 
$\fgas$ is the ICM mass fraction, $\Da(z)=\Ho^{-1}\da(z)$ is the
angular diameter distance, and $T$ is
to be understood as the mean (particle, and not 
emission--weighted) electron temperature; note
that with this understanding, the relation does not
depend on any assumption of isothermality.  
It is this direct relation between observable SZ flux and thermal
gas energy that lies at the heart of some of the advantages of
SZ over X--ray surveys (Bartlett 2001).\\
We have taken care to write the quantities $\fgas$ and $T$ as
general functions of mass and redshift.  In the absence of efficient 
heating/cooling and gas reprocessing, the cluster population
will be fully self--similar.  Simple theoretical
arguments based on energetics during collapse suggest in this case
the existence of a scaling law between cluster 
temperature and mass:
\begin{equation}\label{eq:Tscaling}
T(M,z) = T_*  \left(M_{15} h\right)^{2/3} \left(\Delta (z) E(z)^2\right)^{1/3}
        \left[1-2\frac{\OmL(z)}{\Delta (z)}\right]
\end{equation}
adopting the notation of Pierpaoli et al. (2002).
In this expression, the mass $M_{15}$ is measured in units of 
$10^{15}~$M$_\odot$, $\Delta(z)$ is the full non--linear 
overdensity inside the virial radius relative to the
critical density ($\sim 178$).
In the self--similar model, the gas mass fraction 
$\fgas$ is constant, essentially proportional to the universal ratio
$\OmB/\OmM$.   \\
These scaling expectations are indeed manifest
in hydrodynamical simulations that neglect cooling, and
supported by observations of the more massive clusters with
cooling timescales longer than the Hubble time (Mohr et al. 1999; Allen et al. 2001; Finoguenov et al. 2001; Xu et al. 2001; Arnaud et al. 2002).  Although there is good indication that clusters are
a population with rather regular properties, as suggested by
these scaling laws, there is also direct evidence that it is
not exactly self--similar.  Deviations are most pronounced
for the lower mass objects ($T < 2$~keV), 
as perhaps expected since they
have shorter cooling times and plausible energy injection
mechanisms more readily compete with their gravitational
energy (Ponman et al. 1999; Lloyd--Davies et al. 2000).  
More generally, one may write
\begin{equation}\label{eq:Ygenscaling}
Y(M,z) = Y_{15}(z)M_{15}^{5/3+\alpha}(1+z)^\gamma
\end{equation}
The factor $Y_{15}(z)$ incorporates all the
$z$ dependence of Eqs. (\ref{eq:Yscaling},\ref{eq:Tscaling}) and 
is defined such that the self--similar model corresponds 
to $\alpha=\gamma=0$.

Although we have discussed Eq. (\ref{eq:Ygenscaling}) via
scaling relations for $T$ and $\fgas$, it is  
more pertinent to consider it as a 
single relation between observable SZ flux and 
cluster mass and redshift, one that expresses the 
proportionality between the observable flux and the 
gas thermal energy.
Hydrodynamical simulations incorporating cooling and pre--heating
(da Silva et al. 2003) indicate $\alpha$ between 0.1 and 0.2  
and $\gamma \approx 0$, with a very tight scatter about the
relation, much tighter than corresponding relations for
X--ray quantities.  
This confirms our expectations that the SZ flux
is a more robust quantity than its X--ray 
counterparts.  Even though very low mass clusters
are included in the fit, only mild deviations from
self--similarity in this relation are seen in the simulations.
These deviations are most pronounced
in the low mass systems, while the
higher mass systems appear compatible with the self--similar
scaling laws for $T$--$M$ and $\fgas$.
Caution is still warranted, however, 
due to the unstable nature of cooling, 
issues of numerical resolution, and the fact that these codes
model the gas as a single phase medium; 
the simulation volume also contains mostly low mass systems,
with only a few clusters with $M_{15} > 0.2$.
Simulation results are nonetheless roughly consistent with X--ray 
observations (Borgani et al. 2002; Muanwong et al. 2002). 
 
At our fiducial Planck flux limit of $Y=10^{-4}$~arcmin$^2$, we have 
checked that the counts and value of the angular function at 
30~arcmins are little affected by clusters below $T=2$~keV, 
at least at $\Omega_M < 0.6$.  We therefore concentrate on 
the self--similar case with $\alpha=\gamma=0$.
On the other hand, we consider the normalization
of Eq. (\ref{eq:Ygenscaling}) as a parameter free to
vary within certain limits suggested by X--ray 
observations of $\fgas$ and $T$.  Moscardini et al.
(2002) have argued that this freedom makes it difficult
to use only SZ observations to constrain $\sigma_8$
and $\OmM$.  To overcome this modeling uncertainty,
we combine SZ observations of both the
number counts and the angular correlation function
with constraints arising from the local abundance
of X--ray clusters.  We shall find that
the three kinds of observations are complementary
and lead to constraints on the cosmological parameters.\\
To give a feel for the order of magnitude, we note that
\begin{eqnarray}\label{eq:Ynorm}
\nonumber
Y_{15}(z) & =  \left(7.4\times 10^{-5}h^{7/6}\;{\rm arcmin}^2\right) 
        \left(\frac{T_*}{{\rm keV}}\right)
        \left(\frac{\fgas}{0.07h^{-3/2}}\right) \times \nonumber \\
        &\left(\frac{\Delta (z) E (z)^2}{178}\right)^{1/3}
        \left[1-2\frac{\OmL(z)}{\Delta (z)}\right]
        \frac{1}{\da^2(z)}  \nonumber \\
         & \equiv  Y_*\left(\frac{\Delta (z) E(z)^2}{178}\right)^{1/3}
        \left[1-2\frac{\OmL(z)}{\Delta (z)}\right]
        \frac{1}{\da^2(z)}
\end{eqnarray}
For reference, $T_*=1.2$ according to the simulations of Evrard et al. (1996)
and $\fgas=0.07h^{-1.5}$ from Mohr et al. (1999).  At 
an observation frequency of 2.1~mm, the maximum decrement
of the thermal spectrum, a $Y=7.4 \times 10^{-5}$~arcmin$^2$ 
corresponds to a flux density of $\sim 7$~mJy.
In all the analysis a minimum cluster mass of $10^{14}h^{-1}\;$M$_\odot$ is imposed.

%Figure------------------
\begin{figure*}
\dfig{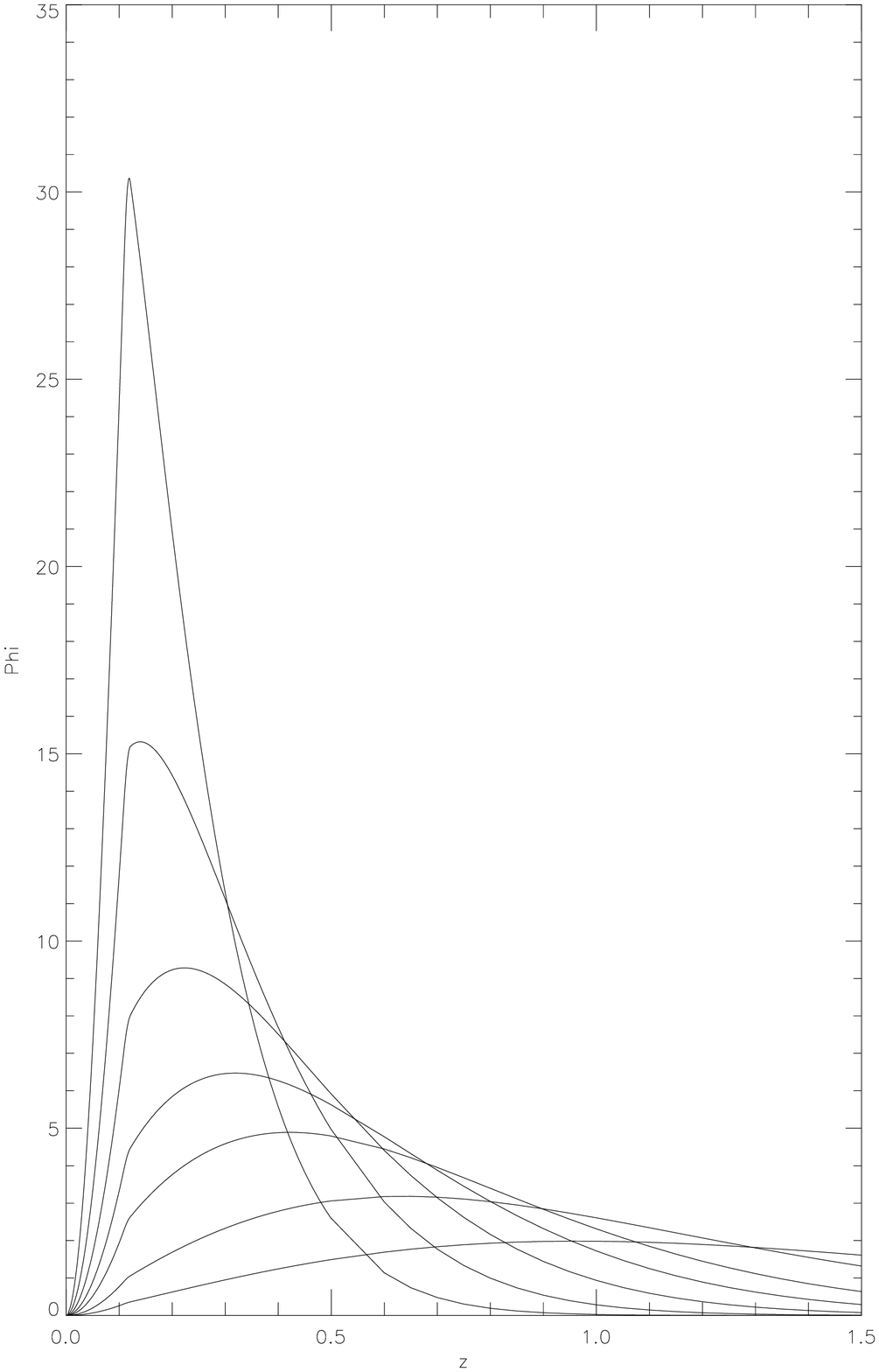}{0.45}{0}{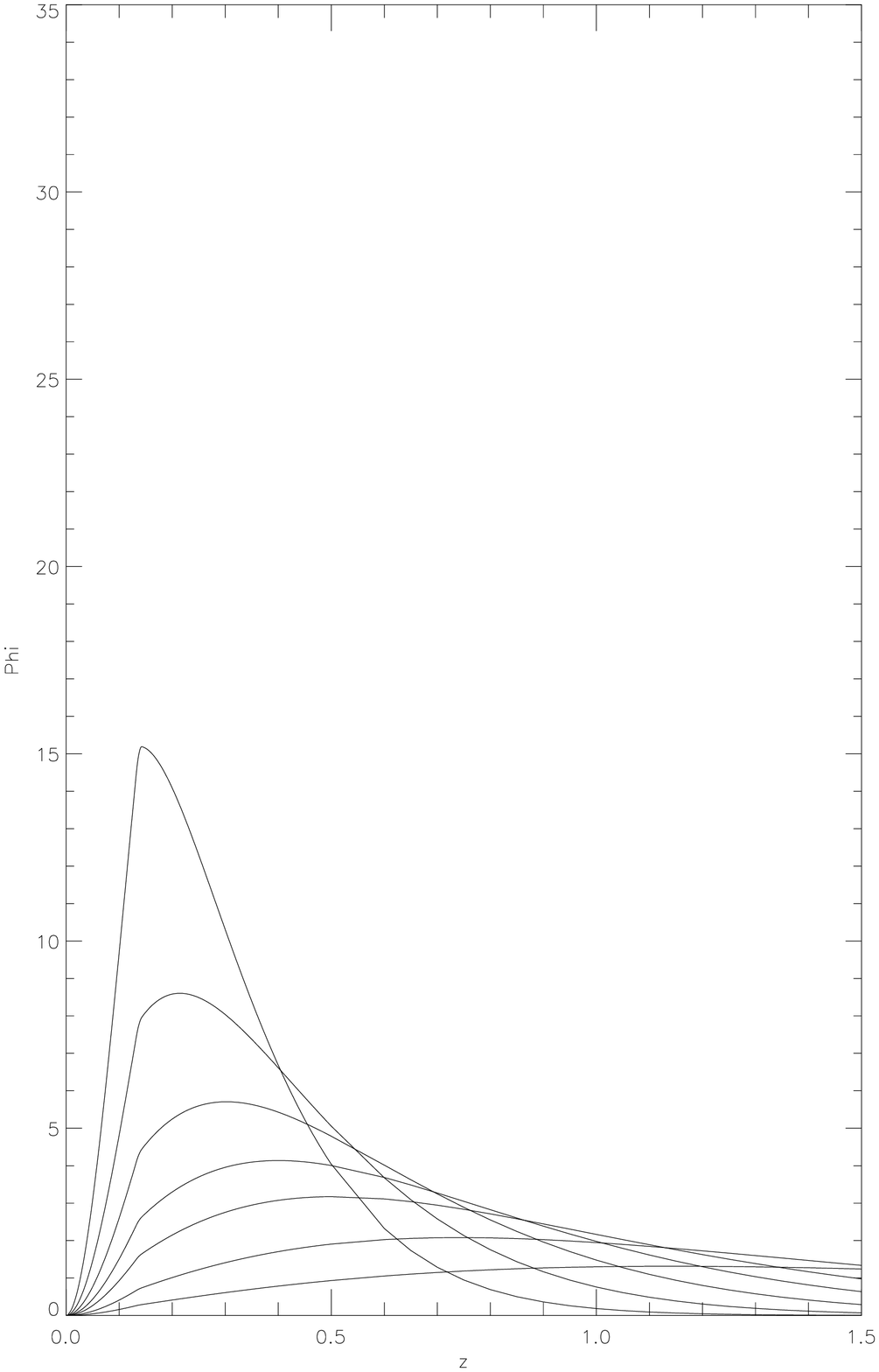}{0.45}{0}
\caption{The selection function $\Phi$ of SZ--detected clusters
with $Y\ge\Ylim=10^{-4}$~arcmin$^2$
(see Eq.~\ref{eq:limber}) as a function of redshift $z$ and 
for different values of $\sigma_8$ in the low--density model
($\OmM=0.3$, on the left) and in the critical model ($\OmM=1$, on 
the right).  From top to bottom, the curves are for increasing 
$\sigma_8=0.5, 0.6, 0.7, 0.8, 0.9, 1.1, 1.4$. The models adopt the same linear matter power spectrum (shape parameter $\Gamma=0.25$) and  $Y_*=7.6\times 10^{-5} h^{7/6}$~arcmin$^2$.}
\label{fig:sel}
\end{figure*}
%----------------------------------------

\section{Results}

\subsection{The angular correlation function}

Calculated angular correlation functions are shown in 
Figure~\ref{fig:models} for three different limiting flux values 
($\Ylim$) in two different flat cosmologies ($\OmM=0.3$ and $\OmL=0.7$,  
with $\sigma_8 = 0.9$, and $\OmM=1$ with $\sigma_8 = 0.7$; e.g.,
Blanchard et al. 2000);
the $Y$--$M$ normalization is $Y_*= 7.6\times 10^{-5} h^{7/6}$~arcmin$^2$.
Our angular correlation function is
consistent with the results of Diaferio et al. (2003)
at a separation of 30~arcmins.  As shown by the latter authors,
small scales are affected by the specific choice of
clustering bias function $b(M,z)$, but the model differences drop to
$\sim 10$~\% at 30~arcmins.  This is comparable to
the statistical measurement errors expected in the
case of the Planck survey, as discussed below.  
We therefore use the angular function at 30~arcmin separation
in our analysis of cosmological parameters. \\
Figure~\ref{fig:sel} shows the cluster selection function $\Phi(z)$,
defined in Eq.~(\ref{eq:limber}),
at $\Ylim=10^{-4}$~arcmin$^2$ in the two cosmologies and for 
various normalizations $\sigma_8$.
The broad break in the angular function in Figure~\ref{fig:models}
corresponds to the break in $\xidm$ just beyond $\sim 10h^{-1}$~Mpc.  
For example, the break occurs $\sim 1\deg$ for 
$\Ylim=10^{-4}$ arcmin$^2$ in the critical model.  According
to Figure~\ref{fig:sel}, the selection function peaks around
$z\sim 0.2$ (for the chosen value of $\sigma_8=0.7$), 
projecting the break to an angular separation 
of $\theta \sim 10h^{-1}/(3000h^{-1}\times 0.2)\sim 1\deg$.
Figures \ref{fig:models} and 
\ref{fig:sel} visually illustrate how the angular function encodes 
information on the catalog radial distribution. 

Although the two models have comparable angular correlations at
the bright end (the upper curves in Fig.~\ref{fig:models} 
corresponding to $\Ylim=10^{-3}$~arcmin$^2$), their dependence
on catalog depth clearly differs.  At the bright end, we are 
mainly observing the local cluster population, which
is essentially the same in both models since the present--day
abundance is the same and the density perturbation power 
spectrum is fixed ($\Gamma=0.25$).  Note that the low--density model
has a slightly steeper slope at small separation, due to
its greater non--linear evolution.  The angular function
of the low--density model decreases and shifts to the left 
more rapidly with survey depth than in the critical model.  
The overall trend is easily understandable and due to
the fact that the selection function broadens and
peaks at higher redshift with survey depth, 
moving correlations to smaller angular scales and
generally washing out the signal as more clusters
are projected along the line--of--sight.
In bright galaxy surveys, which sample the local universe 
where space is approximately Euclidean and galaxy 
evolution may be ignored,
the dependence of the angular function on survey depth 
follows an important {\em scaling law} that is independent of the
underlying cosmological model.  No such universal scaling
law obtains in the SZ case, because the radial
distribution extends out to large redshifts.  For the
relevant flux limits, a SZ catalog therefore samples 
evolution in the cluster population.  Since the cluster population evolves
less rapidly with redshift in the low--density model, the 
angular correlation function therefore shifts down more
rapidly with survey depth than in the case of the critical 
model.  The angular correlation function of SZ clusters is 
therefore a cosmological probe.\\

\subsection{Combining the SZ angular function and counts}

\begin{figure*}
\dfig{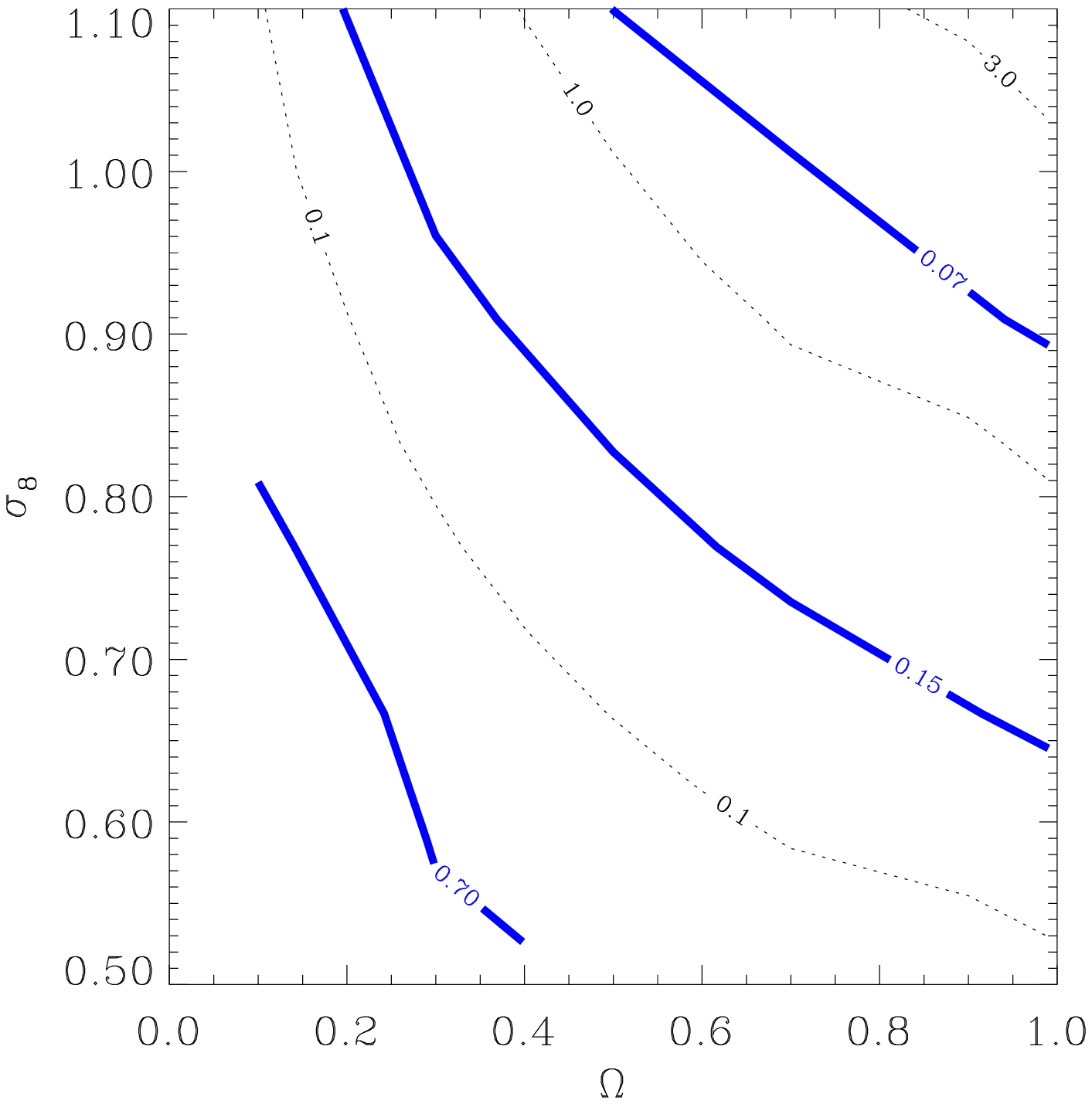}{0.50}{0}{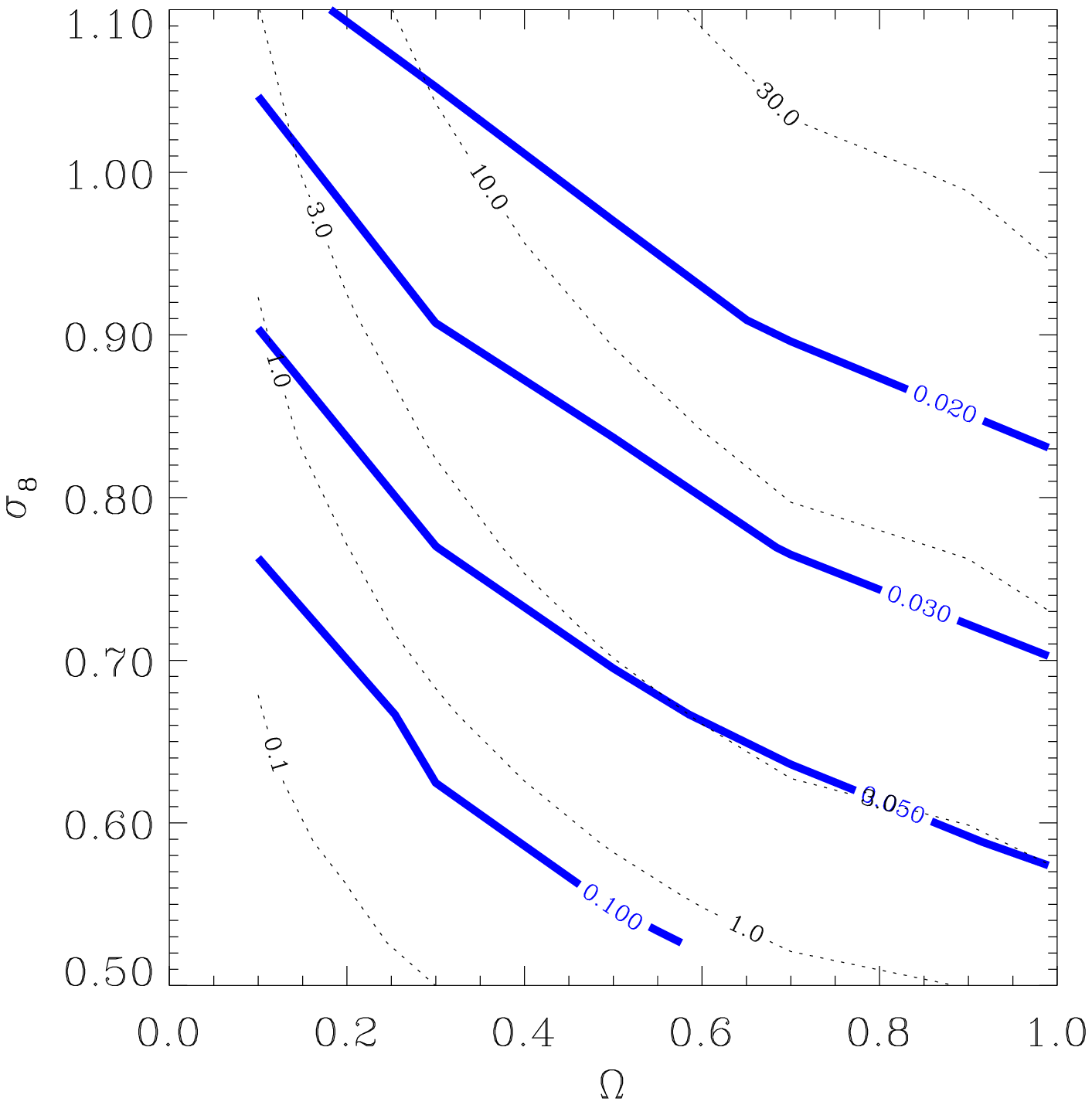}{0.50}{0}
\leftline{\psfig{file=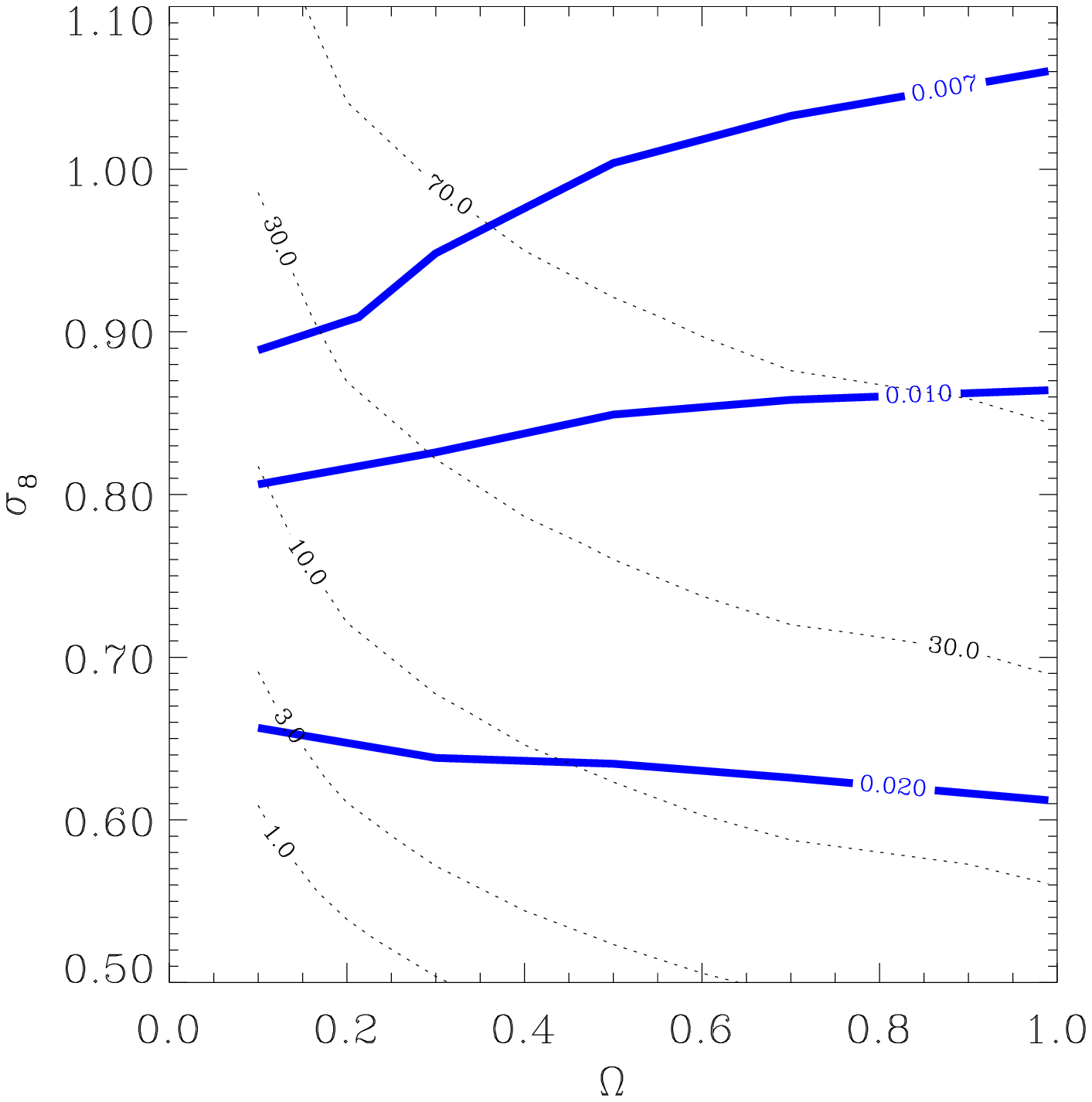,width=9cm}}
\caption{Contours of the SZ angular correlation function are 
shown as solid lines in the $(\sigma_8,\OmM)$--plane for different 
limiting flux values: top left, $\Ylim = 10^{-3}$~arcmin$^2$;  top right, 
$\Ylim = 10^{-4}$~arcmin$^2$; and bottom left, $\Ylim = 10^{-5}$~arcmin$^2$. 
The Hubble constant is fixed at $h=0.7$.
Each contour is labeled with the
value of the correlation function at a separation of $\theta=30$~arcmins.
Contours of the SZ cluster counts at the same flux limits are shown as 
the dotted lines, and labeled in units of 1/sq.deg.  The SZ flux normalization
$Y_*=7.6\times 10^{-5} h^{7/6} $~arcmin$^2$ .}
\label{fig:sigflat}
\end{figure*}

As an illustration of this probe, we next consider the dependence
of the angular function and of the SZ counts on the  
cosmological parameters $(\OmM,\sigma_8)$.  
In Figure~\ref{fig:sigflat} we show the predicted counts and 
angular correlation function for a set of flat models over a grid 
in the $(\sigma_8,\OmM)$--plane and for different limiting flux
values, $\Ylim = 10^{-3}$~arcmin$^2$ (top left),  
$\Ylim = 10^{-4}$~arcmin$^2$ (top right), and $\Ylim = 10^{-5}$~arcmin$^2$ 
(bottom left). Contours of the angular correlation
function are shown as continuous lines and refer to its value
at a separation of $\theta=30$~arcmins.  
The counts contours at the corresponding flux limits are given 
as dotted lines and labeled in units of 1/sq.deg.
This figure is constructed for a self--similar cluster population
with a $Y$--$M$ normalization, $Y_*=7.6\times 10^{-5} h^{7/6}$~arcmin$^2$ 
(see Eq.~\ref{eq:Ynorm}); the
effects of varying ICM properties will be discussed below.\\
We see from the figure that the counts and angular function
carry more complementary information as the catalog
increases in depth.  At the high flux limit of 
$\Ylim=10^{-3}$~arcmin$^2$, the catalog primarily probes
the local universe; the two sets of contours are 
parallel and there is no information permitting us
to constrain the two cosmological parameters.  As we
move deeper, the two sets of contours begin to cross, 
indicating the we are obtaining useful cosmological
information capable of constraining the two parameters.
The limit $\Ylim=10^{-4}$~arcmin$^2$ is representative
of the Planck mission (Aghanim et al. 1997; Bartelmann 2001; 
and references therein), while $\Ylim=10^{-5}$~arcmin$^2$ would
correspond to deeper ground--based surveys (probably
attaining the confusion limit).\\

\begin{figure*}
\dfig{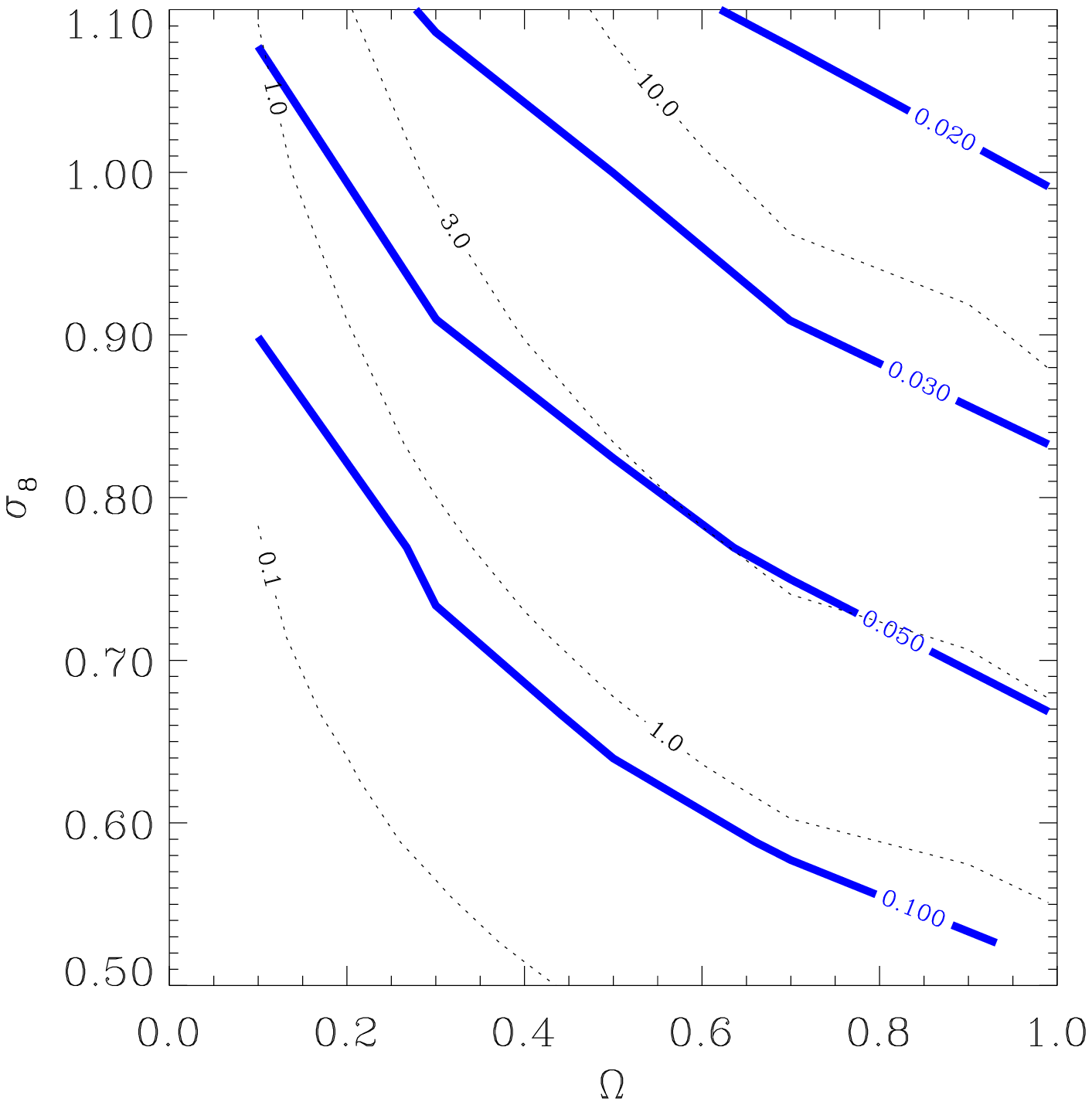}{0.50}{0}{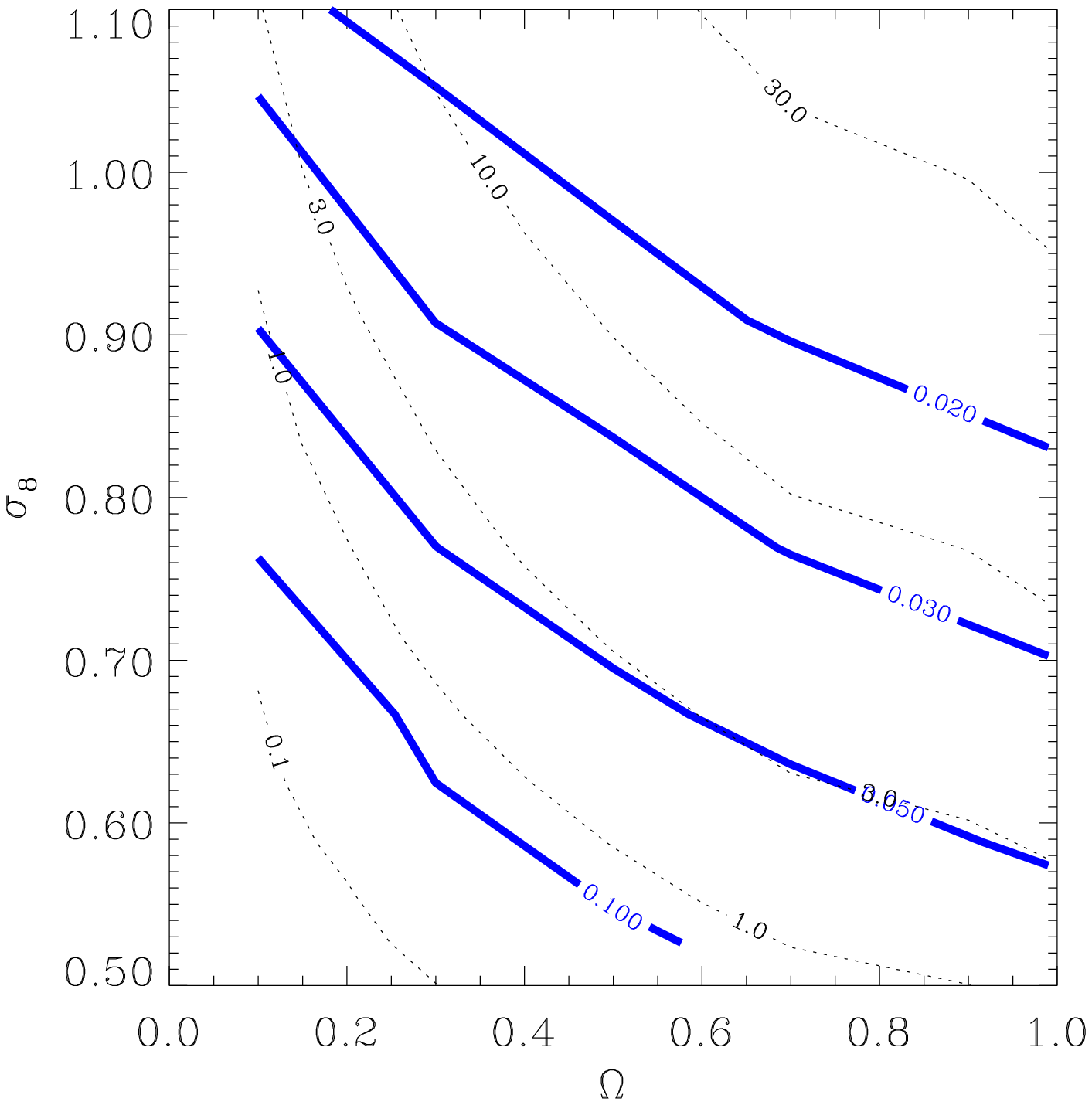}{0.50}{0}
\leftline{\psfig{file=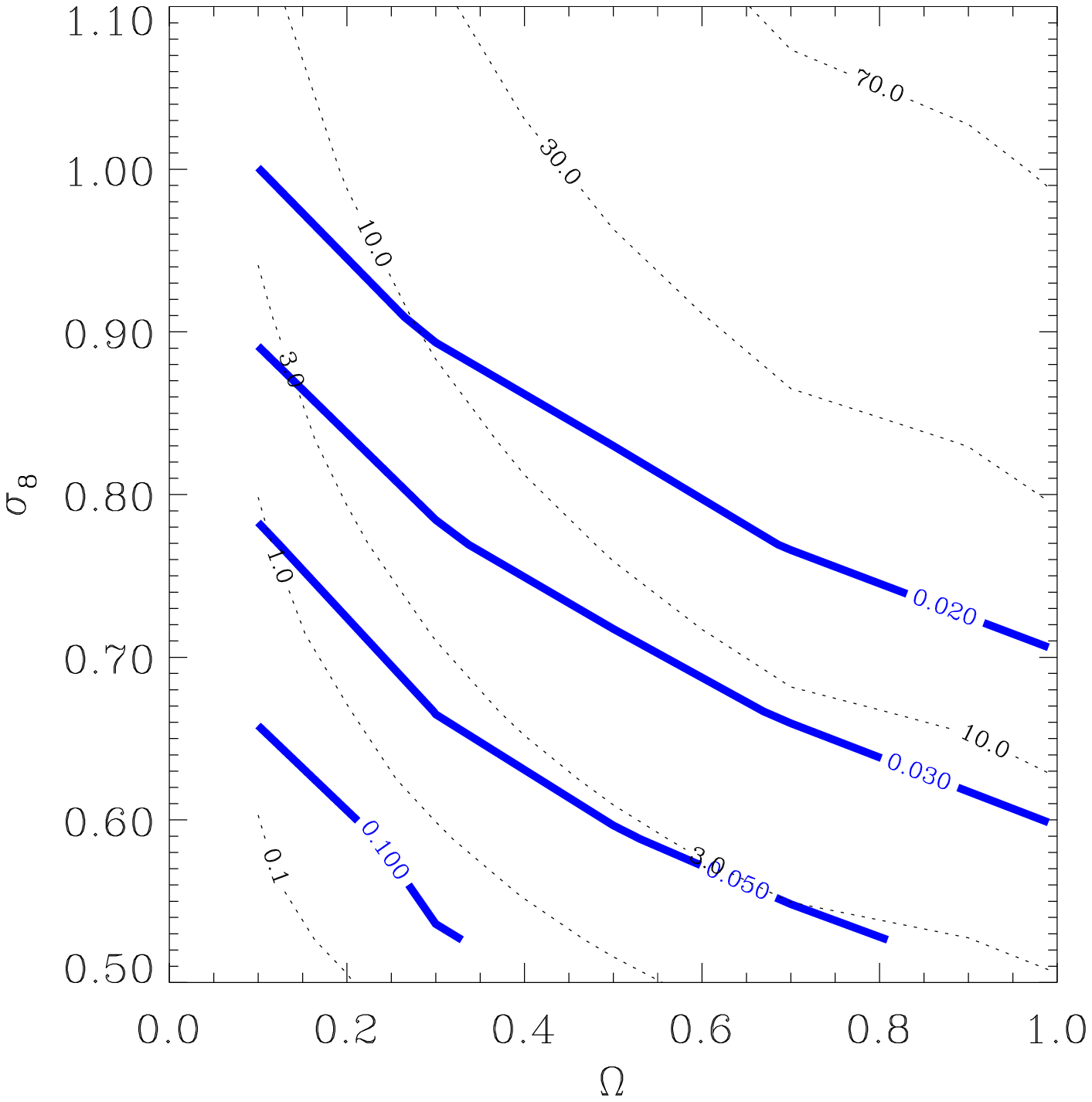,width=9cm}}
\caption{Effect of varying $Y_*$. 
The contours have the same meaning as in the previous figure,
and $h=0.7$ as before.  All three panels correspond to a flux 
limit of $\Ylim=10^{-4}$~arcmin$^2$, but $Y_*$ changes
from $3.8\times 10^{-5} h^{7/6} $~arcmin$^2$ (upper left) to 
$7.6\times 10^{-5} h^{7/6} $~arcmin$^2$ (upper right; 
same as upper right panel of previous figure) and finally
$1.5\times 10^{-4}  h^{7/6}$~arcmin$^2$ (lower left).  
The contours and their spacing are
stretched upward as $Y_*$ decreases.  The intersection
points of the two sets of contours essentially move vertically,
changing in $\sigma_8$ and remaining roughly fixed in $\OmM$.}
\label{fig:sigfB}
\end{figure*}

\subsection{Influence of the ICM}

The single relation Eq.~(\ref{eq:Ygenscaling}) incorporates
all of the ICM physics, and from the nature of the SZ 
flux, we expect it to be rather tight with little scatter.
There is virtually no direct observational information on this 
relation, while numerical simulations (da Silva 
et al. 2003) do indeed confirm a small scatter. They 
also indicate that there is only a slight
deviation from the self--similarity with $\alpha=0.1-0.2$ and
$\gamma=0$ down to masses well below $10^{14}\;$M$_\odot$.
X--ray observations, on the other hand, demonstrate deviations
from self--similarity.  This is most clear for low--mass
systems with temperatures below $T\sim 2$~keV,
but even in the richer systems the observed 
$L$--$T$ deviates from the self--similar
expectation.  These results are beginning to be reproduced
by numerical simulations including cooling/heating (e.g., 
Borgani et al. 2002), including those that indicate only slight 
deviation from self--similarity
for the SZ relation Eq.~(\ref{eq:Ygenscaling}) (Muanwong et al. 2002).  
That X--ray observables show greater deviation
from self--similarity than the SZ flux is not 
surprising, given that the former are more sensitive
to spatial/temperature structure in the gas than
the latter, as emphasized previously.  This is
precisely the advantage of SZ surveys.  

As mentioned, low mass systems with the
greatest deviation from self--similarity in X--rays
contribute little to counts and the angular function
at a flux limit of $\Ylim=10^{-4}$~arcmin$^2$.  
For example, for $\OmM<0.6$ and values of $\sigma_8$ in the 
range 0.6--1, the uncertainty on $\sigma_8$ due to the presence of 
low mass systems with $T<2$~keV is around $5 \%$, well 
within our estimated errors on the angular correlation function 
(Eq. \ref{eq:error} below). 
We therefore
focus on the effects of changing the normalization,
Eq.~(\ref{eq:Ynorm}), of the self--similar
SZ flux--Mass relation ($\alpha=\gamma=0$).
Physically, this represents an uncertainty in the average 
thermal gas energy of galaxy clusters.  
Figure \ref{fig:sigfB} shows the effect of changing 
$Y_*$ at a flux limit of $\Ylim=10^{-4}$~arcmin$^2$.
From the figure, we see that changing this normalization 
stretches the counts 
and angular function contours vertically by roughly the 
same amount, their point of intersection (if any) remaining at 
roughly the same value of $\OmM$. For example, a factor 2 change 
in normalization moves the intersection by $\sim 10-20$~\% 
in $\sigma_8$ (e.g., at $\OmM\sim 0.3$).
This represents an inherent systematic caused by 
modeling uncertainty associated with the ICM.  
Uncertainties related to ICM modeling have been 
extensively discussed by Moscardini et al. (2002) 
for the full 3D spatial correlation function of SZ clusters. \\
Including the uncertain SZ flux normalization, $Y_*$, we now have three
parameters to determine, and so additional information
is needed; for example, an observational program to
determine the $Y$--$M$ relation on a representative
sample of clusters.  Another
tactic is to use the constraint arising from the
local cluster abundance as measured by the X--ray
temperature function.  

\subsubsection{Adding the local cluster abundance constraint}

\begin{figure}
\includegraphics[width=8.5cm]{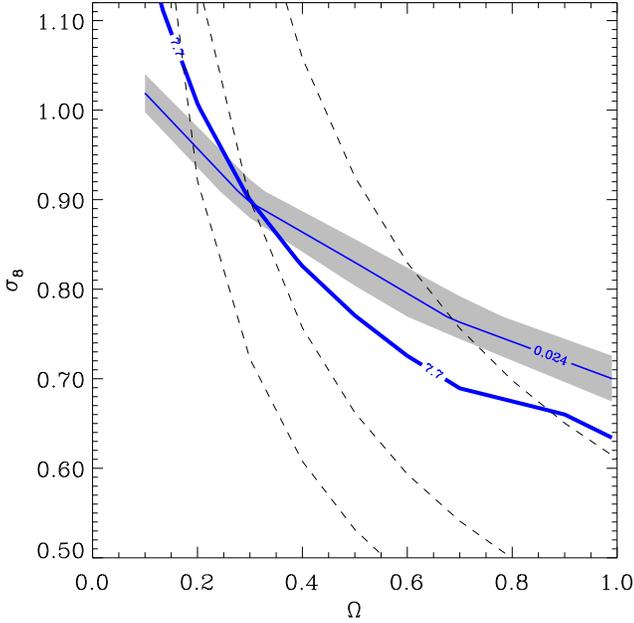}
\caption{Example of constraints in flat models. 
The true underlying model corresponds to $\OmM=0.3$,
$\sigma_8=0.9$, $\fgas=0.07/h^{1.5}$ and
$T_*=1.5$~keV, corresponding to $Y_*=1.1\times 
10^{-4} h^{7/6}$~arcmin$^2$.  For this model, one would observe 
7.7 clusters/sq.deg. for the counts and $w=0.024$ for the angular 
function at 30~arcmins, respectively, both at a flux limit of 
$Y=10^{-4}$~arcmin$^2$.  The bold (heavy blue) line shows 
the constant counts contour in the plane,
while the thin (lighter blue) line indicates the angular 
function contour at the observed value.  The middle 
dashed line represents the constraint from the local
X--ray cluster temperature function, for the true
value of $T_*$; the other two dashed lines correspond
to the constraints for $T_*=2$~keV (bottom dashed curve) and 
$T_*=1$~keV (upper dashed curve).  The three
sets of contours cross at a unique point.
The gray shaded band
indicates the estimated statistical uncertainty on $w$ 
attainable with the Planck survey (the error on the
counts is roughly the thickness of the contour).
}
\label{fig:xabu}
\end{figure}

For a fixed value of $T_*$, the local X--ray temperature
function constrains the cosmological parameters 
$\sigma_8$ and $\OmM$ to a well--defined 
curve in the plane: $\sigma_8 \OmM^{0.6} \approx 0.6T_{*}^{-0.8}$ 
(Pierpaoli et al. 2002).  The exact relation depends somewhat
on the chosen mass function.  
Consider such a constraint, shown as the middle dashed line
in Figure~\ref{fig:xabu}.  Supposing that we measure now both the
SZ counts and angular function, finding 7.7 $~sq.~deg$ and
$w(\theta=30\; {\rm arcmins})=0.024$, respectively, at 
$Y=10^{-4}$~arcmin$^2$ and given as the solid contours in 
the figure.  Apart from the cosmological parameters $\sigma_8$ and
$\OmM$, the SZ contours depend only on $Y_*$, whose
value is uncertain; by changing $Y_*$, we move the SZ contours
around the plane, as previously described.  However, there is
a unique value for $Y_*$ that reduces the intersection of the
three lines to a single point in the plane.  By adjusting $Y_*$ 
to this value, we {\em simultaneously
constrain the cosmological parameters and the normalization $Y_*$}.
In the figure, this intersection lies on top of the true
underlying model.  The determined value of $Y_*$ together with
the known $T_*$ imply a constraint 
on $\fgas$\footnote{We are implicitly assuming that mean electron 
temperature, relevant to the SZ effect, and the observable
X--ray temperature, parametrized by
$T_*$, are the same; this need not be the case.}.
By adding the information on the local cluster abundance, we
have pinned down the relevant ICM physics 
and constrained the cosmological parameters, eliminating
the primary uncertainty in using SZ clustering observations
(Moscardini et al. 2002).
Unfortunately, $T_*$ is at present only poorly known, with
simulations and observations indicating values 
generally in the range $1.2-2$ (Pierpaoli et al. 2002; 
Muanwong et al. 2002; Huterer $\&$ White 2002).  
We expect that it will be much better determined by
the time large SZ catalogs become available, making 
this kind of analysis possible.

This example illustrates the utility of the angular function:  
with only the local cluster abundance and the SZ counts, we 
cannot determine the cosmological parameters $\sigma_8$ and
$\OmM$, due to the uncertainty on $Y_*$.  Adding the angular 
function breaks the degeneracy.  We are thus able to 
constrain the cosmology with the 2D SZ catalog,  
{\em without recourse to redshift determinations}.  
Constraints obtained in this way are particularly 
useful for their complementarity to constraints from 
CMB anisotropy and distance measurements with SNIa.
For example, with SZ clusters we measure $\sigma_8$ directly 
on the relevant scales and constrain $\OmM$, as opposed to the 
physical density $\OmM h^2$ in the case of the CMB.  \\

As mentioned above, clusters below $\sim 2$~keV contribute
little to the contours at our fiducial flux limit of 
$Y=10^{-4}$~arcmin$^2$.  At lower flux limits, more
representative of future ground--based surveys, we could 
expect that the possible effects of deviations from 
self--similar scaling become more important, although simulations
at present indicate only mild effects on the SZ flux.

\subsection{Discussion of statistical errors}

Before concluding, we briefly examine the statistical errors expected
on a measurement of the SZ angular correlation function.
Although a thorough analysis of this issue requires detailed 
simulations of a given survey,  we may nonetheless
make some general arguments to gain insight into what may 
eventually be achieved.  We take the Planck 
survey as our example.
The resolution of the Planck SZ catalog will be on the order of 
5 arcmins (at best), so we can only expect to measure the 
correlations on larger scales (as the $30 \arcmin$ we have chosen in
our analysis), and the  fiducial sensitivity expected is 
$Y\sim 10^{-4}$~arcmin$^2$ (Aghanim et al. 1997; Bartelmann 2001).
Since the angular correlations are small in this context
($\leq 0.1$), we may estimate the (statistical) error on a 
measurement of $w(\theta)$ as the {\em Poisson} variance in the
number of pairs, $\npair$, at this separation (e.g.,
Peebles 1980; Landy \& Szalay 1993).  This quantity is determined by the counts as follows:
Suppose that we measure $w$ in a annular bin of width $\Dtheta$ 
at angular distance $\theta$ from a cluster.  
The mean number of clusters in this ring is $\langle n\rangle = 
2\pi\theta\Dtheta \Sigma(\Ylim)$, from which we deduce that the total
number of pairs at this separation in a catalog of $N$ clusters
is about $\npair \approx (1/2) N\times \langle n\rangle = 
N \pi\theta \Dtheta\Sigma$.  In other words, we estimate the
error to be
\begin{eqnarray}
\label{eq:error}
\Delta w(\theta,\Ylim) &\approx \frac{1}{2\pi}
        \left[\Sigma(\Ylim)\right]^{-1}\frac{1}{\theta}
        \left(\frac{\Dtheta}{\theta}\right)^{-1/2} \nonumber \\
&\approx
        2.78\times 10^{-3}\left(\frac{{\rm deg}}{\theta}\right)
        \left(\frac{\Dtheta}{\theta}\right)^{-1/2}
        \left(\frac{\Sigma}{{\rm deg}^2}\right)^{-1}
\end{eqnarray}
Notice that this statistical error depends essentially 
on the number counts $\Sigma$.  This leads to the gray shaded
area in Figure \ref{fig:xabu}

\section{Discussion and Conclusion}

        We have calculated the angular correlation function of 
SZ--detected clusters in order to evaluate its usefulness for 
extracting cosmological information directly from a 2D SZ cluster 
survey, before 3D follow--up.  The angular correlation function
of SZ detected clusters differs from the angular power spectrum 
of SZ induced CMB anisotropies, which is dominated by the Poisson
term.  The different scaling of angular correlations with survey 
depth visually demonstrates the cosmological sensitivity of the 
angular function (See Figs~\ref{fig:models},~\ref{fig:sel}).
As illustration, we considered the parameter pair 
$(\OmM,\sigma_8)$ in the context of flat $\Lambda$--CDM models.  
We found that at sufficient depth (e.g., $\Ylim 
\sim 10^{-4}$~arcmin$^2$, comparable to the Planck mission), 
the counts and angular function combined can constrain
these parameters.  Modeling uncertainty associated
with the ICM may be reduced by measuring the $T$--$M$ relation
and adding the corresponding constraint from the local abundance 
of X--ray clusters; in this way, the two cosmological
parameters and the SZ flux normalization $Y_*$ may be found.
Deeper ground--based surveys (e.g., $\Ylim\sim 10^{-5}$~arcmin$^2$), 
will pick up a larger number of low--mass objects whose 
ICM properties require more careful modeling of deviations
from self--similarity.

The accuracy and precision with which one will be able to 
measure the angular function (and the counts) is clearly
an important issue.  We only briefly touched on this point
with a simple estimate of the expected statistical errors 
expected on $w(\theta)$ in the case of the Planck mission.
A more detailed examination incorporating simulations of 
the SZ survey characteristics, including instrument noise and 
foreground contamination, as well as of the catalog extraction 
algorithms, is needed.  
The survey selection function will in reality
depend on these details (Bartlett 2001; Melin et al. 2003; White 2003),
and so will the errors on the counts and measured angular
function.   

        It is clear that the most powerful constraints from an SZ survey
will come from its measured redshift distribution.  
The interesting point is, however, that the method proposed here 
increases the immediate scientific return from an SZ survey 
by offering a way to obtain pertinent cosmological
constraints using only a 2D SZ catalog, without recourse
to follow--up observations.

\begin{acknowledgements}
We are grateful to Lauro Moscardini, Antonaldo Diaferio and Antonio da Silva,
and the anonymous referee for their very useful comments that helped the 
improvement of this paper.
We would also like to thank Antonella De Luca and Milan Maksimovic, 
who first opened the contacts between us.
S. Mei acknowledges support from the European Space Agency External 
Fellowship programme.

\end{acknowledgements}


\begin{thebibliography}{}


\bibitem[Aghanim  et al. 1997]{agh97}
Aghanim, N., de Luca, A., Bouchet, F.R. {\it et al.} 1997, A\&A, 325, 9


\bibitem[Aghanim  et al. 2002a]{agh02a}
Aghanim, N., Castro, P.G., Melchiorri, A. {\it et al.} 2002a, A\&A, 393,
381

\bibitem[Aghanim N. et al. 2002b]{agh02b}
Aghanim, N., Hansen  S., Pastor S. {\it et al.} 2002b, astro-ph/0212392

\bibitem[Allen et al. 2001]{all01}
Allen, S. W., Schmidt, R.W., Fabian A.C., 2001, MNRAS, 328, L37

\bibitem[Arnaud et al. 2002]{arn02}
Arnaud, M., Aghanim, N., Neumann, D. M. 2002, 389, 1

\bibitem[Barbosa et al. 1996]{bar96}
Barbosa D., Bartlett J.G., Blanchard A. {\it et al.}, 1996, A\&A, 314, 13

\bibitem[Bardeen et al. 1986]{bar86}
Bardeen, J.M., Bond, J.R., Kaiser, N., Szalay, A.S. (BBKS) 1986, ApJ, 304,15

\bibitem[Bartelmann 2001]{bart01}
Bartelmann, M. 2001,  A\&A, 370, 754

\bibitem[Bartelmann \& White 2002]{bart02}
Bartelmann, M. \& White, S.D.M. 2002, A\&A, 388, 732


\bibitem[Bartlett 2000]{bar00}
Bartlett J.G. 2000, astro-ph/0001267 

\bibitem[Bartlett 2001]{bar01}
Bartlett J.G. 2001, review in "Tracing cosmic evolution with galaxy clusters" (Sesto Pusteria 3-6 July 2001), ASP Conference Series in press ,astro-ph/0111211


\bibitem[Benson et al. 2002]{ben02}
Benson, A.J., Reichardt, C., Kamionkowski, M. 2002, MNRAS, 331, 71

\bibitem[Benson et al. 2003]{ben03}
Benson, A.J., Church, S.E., Ade, P. A. R. {\it et al.} 2003, astro-ph/0303510


\bibitem[Birkinshaw 1999]{bir99}
Birkinshaw M. 1999, Proc. 3K Cosmology, 476, American Institute of
Physics, Woodbury, p. 298

\bibitem[Blanchard et al. 2000]{blanchard00}
Blanchard A., Sadat R., Bartlett J.G. et al. 2000, A\&A 362, 809

\bibitem[Bond et al. 2002]{bon02}
Bond J.R., Contaldi C.R., Pen, U-L. et al. 2002, submitted to ApJ, astro-ph/0205386 

\bibitem[Borgani et al. 2002]{borgani02}
Borgani S., Governato F., Wadsley et al. 2002, MNRAS 336, 409

\bibitem[Carlstrom et al. 2002]{car02}
Carlstrom, J.E., Holder, G.P., Reese, E.D. 2002, ARA\&A, 40, 643

\bibitem[Carroll et al. 1992]{car92}
Carroll, S.M.; Press, W.H., Turner, E. L. 1992, ARA\&A, 30, 499

\bibitem[Colafrancesco et al. 1997]{col97}
Colafrancesco, S., Mazzotta, P., Vittorio, N. 1997, ApJ, 488, 566



\bibitem[da Silva et al. 2003]{das03}
da Silva,  {\it et al.} 2003, in preparation

\bibitem[Diaferio et al. 2003]{dia03}
Diaferio, A., Nusser, A., Yoshida, N. {\it et al.} 2003, MNRAS, 338, 433

\bibitem[Eke et al. 1996]{eke96}
Eke, V.R., Cole, S., Frenk, C.S. 1996, MNRAS, 282, 263

\bibitem[Evrard et al 1996]{ev96} Evrard, A.E., Metzler, C.A. \&
Navarro, J.F. 1996, ApJ 469, 494


\bibitem[Finoguenov 2001]{Fino2001} 
Finoguenov, A,  Reiprich, T.H. \&  B{\"o}hringer, H. 2001, A\&A 368, 749

\bibitem[Grainge et al. 2002]{vsa02} 
Grainge K., Carreira, P., Cleary K. {\it et al.} 2002,  astro-ph/0212495

\bibitem[Grego et al. 2002]{gre02}
Grego, L., Carlstrom, J.E.,  Reese, E.D. {\it et al.} 2002, \apj, 552, 2

\bibitem[Haiman et al. 2001]{hai01}
Haiman, Z., Mohr, J.J., Holder, G. 2001, ApJ, 553, 545

\bibitem[Holder et al. 2001]{hol01} 
Holder, G., Haiman, Z., Mohr, J.J. 2001, ApJ, 560, L111

\bibitem[Holder G.P. 2002]{hol02} 
Holder G.P. 2002,\apj, 578, L1

\bibitem[Holzapfel et al. 1997]{hol97}
Holzapfel, W.L., Ade, P.A.R., Church, S.E.  {\it et al.} 1997, \apj,
481, 35

\bibitem[Huterer \& White 2002]{hun02}
Huterer, D. \& White, M. 2002, ApJL, 578, L95

\bibitem[Jenkins et al. 2001]{jen01}
Jenkins A., Frenk C.S., White S.D.M. {\it et al.} 2001, MNRAS, 321, 372

\bibitem[Jones et al. 2002]{jon02} 
Jones M.E., Edge A.C., Grainge, K. {\it et al.} 2001, astro--ph/0103046

\bibitem[Kneissl et al. 2001]{kne01}
Kneissl, R., Jones, M.E., Saunders, R. {\it et al.} 2001, MNRAS, 328, 783 

\bibitem[Komatsu \& Kitayama 1999]{kom99}
Komatsu, E., Kitayama, T. 1999, ApJ, 526, L1

\bibitem[Komatsu \& Seljak 2002]{kom02}
Komatsu, E. \& Seljak, U. 2002, MNRAS, 336, 1256

\bibitem[Landy \& Szalay 1993]{lan93}
Landy, S.D. \& Szalay, A.S.  1993, ApJ, 412, 64

\bibitem[Limber 1953]{limber53} Limber D.N. 1953, ApJ 117, 134

\bibitem[Lloyd--Davies 2000]{lloyddavis00} Lloyd--Davis
E.J., Ponman T.J. \& Cannon D.B. 2000, MNRAS 315, 689

\bibitem[Mason et al. 2001]{mason01} 
Mason B.S., Myers S.T. \& Readhead A.C.S. 2001, ApJ 555, L11

\bibitem[Melin 2003]{melin03} 
Melin, J.--B.  {\it et al.} 2003, in preparation


\bibitem[Mo \& White 1996]{mo96}
Mo, H.J., White, S.D.M. 1996, MNRAS, 282, 347

\bibitem[Mo \& White 2002]{mo02}
Mo, H.J., White, S.D.M. 2002, MNRAS, 336, 112

\bibitem[Mohr et al. 1999]{Mohr99} 
 Mohr, J.J., Mathiesen, B.,   Evrard, A. E.
1999, ApJ 517, 627

\bibitem[Moscardini et al. 2002]{mos02}
Moscardini, L., Bartelmann, M., Matarrese, S. {\it et al.} 2002, MNRAS, 340, 102

\bibitem[Muanwong et al. 2002]{mua02}
Muanwong, O., Thomas P.A., Kay, S.T. {\it et al.} 2002, MNRAS, 336, 527


\bibitem[Oh et al. 2003]{oh03}
Oh, S.P., Cooray, A., Kamionkowski, M. 2003, astro-ph/0303007


\bibitem[Peacock \& Dodds 1996]{pea96}
Peacock J. A., Dodds, S. J. 1996, MNRAS, 280, L19

\bibitem[Peebles 1980]{peebles80} Peebles P.J.E. 1980, The Large--Scale
Structure of the Universe, Princeton Series in Physics, Princeton University
Press (Princeton: New Jersey) 


\bibitem[Peebles 1993]{peebles93} Peebles P.J.E. 1993, Principles of
Physical Cosmology, Princeton Series in Physics, Princeton University
Press (Princeton: New Jersey) 

\bibitem[Percival et al. 2001]{percival01}
Percival  W.J., Baugh C.M., Bland--Hawthorn J. {\it et al.}
2001, MNRAS 327, 1297
\bibitem[Pierpaoli et al. 2002]{pie02}
Pierpaoli E., Borgani S., Scott D. et al. 2002, astro-ph/0210567

\bibitem[Ponman et al. 1999]{ponman99} Ponman T.J., Cannon D.B.,
        Navarro J.F. 1999, Nature 397, 135

\bibitem[Reed et al. 2003]{ree03}
Reed, D., Gardner J., Quinn T. {\it et al.} 2003, astro-ph/0301270 

\bibitem[Reese et al. 2002]{rie02}
Reese, E.D., Carlstrom, J.E., Joy, M. {\it et al.} 2002,\apj, 581, 53


\bibitem[Sheth \& Tormen]{she99}
Sheth R.K. \& Tormen G. 1999, MNRAS, 308, 199

\bibitem[Sheth et al. 2001]{she01}
Sheth R.K., Mo H.J. \& Tormen G., 2001, MNRAS, 323, 1

\bibitem[Spergel et al. 2003]{spe03}
Spergel, D.N., Verde L., Peiris H.V. {\it et al.} 2003, astro-ph/0302209

\bibitem[Sunyaev \& Zeldovich 1970]{sun70}
Sunyaev, R. A. \&  Zel'dovich, Ya. B. 1970, Comments Astrophys. Space
Phys., 2, 66

\bibitem[Sunyaev \& Zeldovich 1972]{sun72}
Sunyaev, R. A. \&  Zel'dovich, Ya. B. 1972, Comments Astrophys. Space
Phys., 4, 173

\bibitem[Weller et al. 2002]{wel02}
Weller, J., Battye, R.A., Kneissl, R. 2002, PhRvL, 88, 1301

\bibitem[White 2003]{whi03}
White M. 2003, astro--ph/0302371

\bibitem[Xu2001]{xu2001} Xu, H., Jin, G.,Wu, X.--P. 2001, ApJ 553, 78

\end{thebibliography}
\end{document}